\documentclass{emulateapj}
\usepackage{apjfonts}

\newcommand{\igr}{\hbox{IGR J2018+4043}}

\newcommand{\sw}{{\sl Swift}}
\newcommand{\xrt}{{XRT}}
\newcommand{\il}{{\sl INTEGRAL}}
\newcommand{\ii}{{ ISGRI}}
\newcommand{\gc}{\hbox{$\gamma$-Cygni}}

\def\beq{\begin{equation}}
\def\enq{\end{equation}}
\def\begar{\begin{eqnarray}}
\def\endar{\end{eqnarray}}

\def\lsim{\;\raise0.3ex\hbox{$<$\kern-0.75em\raise-1.1ex\hbox{$\sim$}}\;}
\def\gsim{\;\raise0.3ex\hbox{$>$\kern-0.75em\raise-1.1ex\hbox{$\sim$}}\;}
\def \gr{$\gamma$-ray }

\def\beq{\begin{equation}}
\def\enq{\end{equation}}
\def\begar{\begin{eqnarray}}
\def\endar{\end{eqnarray}}
\def\mathnew{\mathsurround=0pt}
\def\simov#1#2{\lower .5pt\vbox{\baselineskip0pt \lineskip-.5pt
        \ialign{$\mathnew#1\hfil##\hfil$\crcr#2\crcr\sim\crcr}}}

\def\etal{{ et al. }}

\def\enf{\rm ~ergs~cm^{-2}~s^{-1}}
\def\fl{\rm ~cm^{-2}~s^{-1}}

\def \chan {{\sl Chandra}}
\def \swift {{\sl Swift}}
\def \xmm {{\sl XMM-Newton}}
\def \isgri {ISGRI}

\def\igr{IGR~J2018+4043~}
\def\gc{$\gamma$-Cygni~}
\def \degmark{^\circ}

\def \hcm {\hbox {\ifmmode $ atom cm$^{-2}\else atom cm$^{-2}$\fi}}
\def \arcmin {\hbox{$^\prime$} }
\def \arcsec {\hbox{$^{\prime\prime}$} }

\def\approxgt{\mathrel{\hbox{\rlap{\lower.55ex \hbox {$\sim$}}
        \kern-.3em \raise.4ex \hbox{$>$}}}}
\def\approxlt{\mathrel{\hbox{\rlap{\lower.55ex \hbox {$\sim$}}
        \kern-.3em \raise.4ex \hbox{$<$}}}}
\def\mathnew{\mathsurround=0pt}

\shorttitle{Hard X-ray source IGR J2018+4043}
\shortauthors{A.M.Bykov et al.}

\begin{document}
\title{
On the nature of the hard X-ray source IGR~J2018+4043}

\author{A.M.\ Bykov\altaffilmark{1}, A.M.\ Krassilchtchikov\altaffilmark{1}, Yu.A.\ Uvarov\altaffilmark{1},
J.A.\ Kennea\altaffilmark{2}, G.G.\ Pavlov\altaffilmark{2}, G.M.\
Dubner\altaffilmark{3}, E.B.\ Giacani\altaffilmark{3},
H.~Bloemen\altaffilmark{4}, W.\ Hermsen\altaffilmark{4,5},
J.\ Kaastra\altaffilmark{4}, F.\ Lebrun\altaffilmark{6},
M.\ Renaud\altaffilmark{6}, R.\ Terrier\altaffilmark{7},
M.\ DeBecker\altaffilmark{8}, G.\ Rauw\altaffilmark{8},
J.-P.\ Swings\altaffilmark{8}} \altaffiltext{1}{A.F.\ Ioffe Institute
of Physics and Technology, St.\ Petersburg, Russia, 194021;
byk@astro.ioffe.ru} \altaffiltext{2}{Pennsylvania State University,
525 Davey Laboratory, University Park, PA 16802}
\altaffiltext{3}{Instituto de Astronom\'ia y F\'isica del Espacio
(IAFE), CC 67, Suc. 28,
    1428 Buenos Aires, Argentina}
\altaffiltext{4}{SRON Netherlands Institute for Space Research,
Sorbonnelaan 2, 3584 CA Utrecht, The Netherlands}
\altaffiltext{5}{Astronomical Institute ``Anton Pannekoek'',
University of Amsterdam, Kruislaan 403,
             NL-1098 SJ Amsterdam, The Netherlands}
\altaffiltext{6}{CEA-Saclay, DSM/DAPNIA/Service d'Astrophysique,
91191 Gif-sur-Yvette Cedex, France} \altaffiltext{7}{APC-UMR 7164,
11 Place M.Berthelot, 75231
            Paris, France}
\altaffiltext{8}{Institut d'Astrophysique et de G\'eophysique,
Universit\'e de Li\`ege,
       All\'ee du 6 Ao\^ut 17, B\^at B5c, 4000 Li\`ege, Belgium }

\begin{abstract}
We found a very likely counterpart to the recently discovered hard
X-ray source \igr\ in the multi-wavelength observations of the
source field. The source, originally discovered in the 20-40 keV
band, is now confidently detected
also in the 40-80 keV
band, with a
flux of $(1.4 \pm 0.4) \times 10^{-11} \enf$.
 A 5 ks \swift\
observation of the \igr\
field revealed a hard point-like source
with the observed 0.5--10 keV flux of
3.4$^{+0.7}_{-0.8}\times$10$^{-12}$ erg cm$^{-2}$ s$^{-1}$
(90\% confidence level)
 at $\alpha = 20^{\rm
h}18^{\rm m}38\fs 55$, $\delta = +40^\circ 41' 00\farcs4$ (with a
4\farcs2 uncertainty). The combined {\sl Swift}-{\sl Integral}
spectrum can be described by an absorbed power-law model with photon
index $\Gamma=1.3\pm 0.2$ and $N_{\rm H}=6.1^{+3.2}_{-2.2}\times
10^{22}$ cm$^{-2}$. In archival optical and infrared data we found a
slightly extended and highly absorbed object at the \swift\ source
position. There is also an extended VLA 1.4 GHz source peaked at a
beam-width distance
 from the optical and X-ray positions. The observed
morphology and multiwavelength spectra of \igr\ are consistent with
those expected for an obscured accreting object, i.e. an AGN or a
Galactic X-ray binary. The identification suggests possible
connection of \igr\ to the bright $\gamma$-ray source GEV J2020+4023
detected by {\sl COS B} and {\sl CGRO} EGRET.
\end{abstract}

\keywords{ISM: individual (\gc)---X-rays: individual
(IGR~J2018+4043; 3EG\,J2020+4017)} 


\section{Introduction}
The first hard X-ray imaging observation of the Cygnus region with
the IBIS-ISGRI imager (Lebrun et al.\ 2003) aboard the {\sl
International Gamma-Ray Astrophysics Laboratory} ({\sl INTEGRAL};
Winkler et al.\ 2003) has revealed a new source, IGR J2018+4043
(Bykov et al.\ 2004). The 320 ks fully coded field of view (FCFOV)
observation yielded the source  flux of $(1.7 \pm 0.4) \times
10^{-11} \enf$ in the 25-40 keV band. The Cygnus region is very rich
in starforming sites, X-ray binaries, and supernova remnants (SNRs).
\igr\ is located in the field of the SNR G78.2+2.1
(commonly known as $\gamma$-Cygni SNR).
A high-energy \gr source, 2CG 078+2, in the \gc\ field was
discovered with the {\sl COS B} satellite (e.g., Swanenburg et al.\
1981). Observations with {\sl CGRO} EGRET showed that this
source (2EG\,J2020+4026 = 3EG\,J2020+4017 = GEV\,J2020+4023), whose
photon flux is $\sim$ 1.2 $\times 10^{-6}\fl$ above 100 MeV (Sturner
\& Dermer 1995; Hartman et al.\ 1999), is the brightest among
apparently steady unidentified EGRET sources.
{\sl
Whipple} \gr\ telescope observations (e.g., Buckley et al.\ 1998)
have established an upper limit of 2.2 $\times 10^{-11}\fl$ for the
flux above 300 GeV, indicating a break in the spectrum above a few
GeV. Bykov et al.\ (2004) suggested that the EGRET and \isgri\
sources could be the $\gamma$-ray and hard X-ray counterparts of the
same source in \gc, e.g.,
a Geminga-like pulsar or
extended emission from particles accelerated
in the supernova ejecta/shell.
In particular, the interaction of the wind of the early type O9V
star HD 193322, located just 7\arcmin\ from IGR J2018+4043, with
\gc\ would be a plausible mechanism
of particle acceleration.
On the other hand, the \isgri\ source position
within the SNR field does not
necessarily imply their physical connection. An
example of such a confusion is the source IGR J17204--3554
projected onto the rich Galactic starforming region NGC 6334. The
most likely identification of that source is a highly obscured
extragalactic object NGC 6334B (see Bassani et al.\ 2005;
Bykov et al.\ 2006).

Given the nominal 12\arcmin\ angular resolution (FWHM) and about
1\arcmin point source localization of \isgri\ (Lebrun et al.\ 2003), a
reliable identification of the relatively weak (4--5 $\sigma$) \ii\
source requires high-resolution
observations with other instruments.
We report here
the results of recent 1.5 Ms \hbox{FCFOV} observations of the
source with \isgri\ (\S2.1) and
the detection of a counterpart of
 IGR J2018+4043 in the 0.5--10
keV band with {\sl Swift} XRT (\S2.2). The accurate positioning of
the {\sl Swift} source allowed us  to find its possible
radio/IR/optical counterparts in archival data (\S2.3 and \S2.4).
Based on these multiwavelength observations, we suggest that the
source is a strongly obscured accreting source, likely an AGN, and
discuss the possibility that the EGRET source is indeed its
$\gamma$-ray counterpart (\S3).

\section{Observations and Data Analysis}

\begin{figure*}
\centering
\includegraphics[width=0.8\textwidth,angle=0]{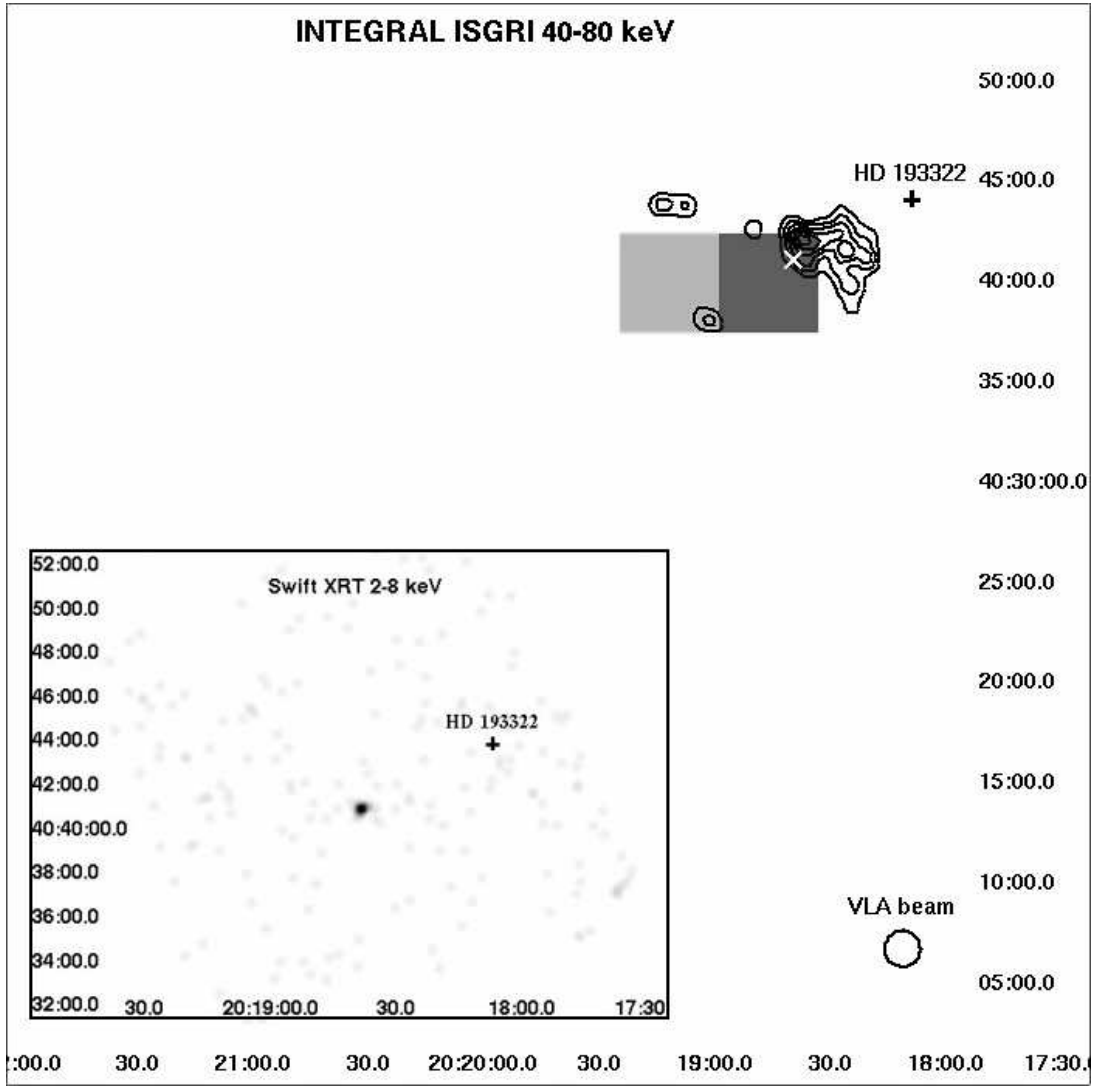}
\caption{ 40-80 keV ISGRI significance map ($\geq 3.5\,\sigma$) of
the \gc\ field, with VLA 1.4 GHz contours overlaid. The light and
dark $5'\times 5'$ ISGRI pixels correspond to 3.6 and 3.8$\,\sigma$
significance, respectively. The VLA contours run from 9 to 73
mJy/beam on a linear scale.
 The \sw\ source position is marked by the white cross.
The smoothed \sw\ \xrt\ 2-8 keV
image is shown in the inset.
The bright O9V star HD193322 was removed from the \xrt\ image.
}
\label{fig1}
\end{figure*}

\subsection{\isgri\ observation}
The field of \gc\ has been frequently observed with \il\ ISGRI.
We used 1.5 Ms of \ii\ good time archival FCFOV data  taken
at revolutions 19--255 (2002 Dec 9 -- 2005 Nov 16). We applied the
selection criteria taking only the science windows
where the
count rate was within $\pm 1\, \sigma$
around the mean value determined by
the whole 1.5 Ms exposure.
The standard reduction was performed with the OSA 5.1 software
 \citep{osa}. Pixel spreading was switched off for a more accurate flux
estimation. Broad energy bands were used to increase S/N. The
source was detected
with a flux of (1.2$\pm$0.2) $\times$10$^{-11}$ erg cm$^{-2}$
s$^{-1}$ in the 20-40 keV band, and (1.4$\pm$0.4)$\times$10$^{-11}$
erg cm$^{-2}$ s$^{-1}$ in the 40-80 keV band, obtained by cross
calibration with the Crab (see, e.g., Lubinski et al.\ 2004).
The flux in 20- 40 keV band is below that quoted in
Bykov \etal (2004) due to the new background model used in OSA 5 and
more conservative selection criteria that we applied now to the much
longer exposure dataset available.  The 40-80 keV \ii\ significance
map is presented in Figure {\ref{fig1}}.

To search for \isgri\ flux variations, we grouped all events in
three time bins
(the maximum number allowed by the
count statistics), but found no
significant flux variations.
With the scarce statistics available, we can only
exclude flux variations
greater than a factor of 4 on a timescale
of about a year.

\subsection{Swift observation}
The field of \igr\ was observed with \sw\ \xrt\ \citep{burrows} on
2006 March 26, 29, and 30, with a total exposure of 4.8 ks
\citep{kennea}. The \xrt\ was operated in photon counting mode. The
data were reduced with the standard HEAsoft package (ver.\ 6.0.5).
The event files were processed with the {\tt xselect} tool
(grades
0--12 were used).
The XSPEC package (ver.\ 12.1.1)
was used for the spectral analysis.
A point-like source was detected in the XRT image
at a position consistent with that of the ISGRI source
 (see
Fig.\ \ref{fig1}):
$\alpha$=20$^{\rm h}$18$^{\rm m}$38$\fs$55,
$\delta$=+40$^\circ$41$'$00$\farcs$4 (J2000), with a position uncertainty
of 4$\farcs$2
(errors here and below in this subsection are at the 90\% confidence level).

We extracted 105 source counts from an aperture of $47\farcs$2
radius and estimated the observed 0.5--10 keV flux
$F_X=3.4^{+0.7}_{-0.8}\times 10^{-12}$ erg cm$^{-2}$ s$^{-1}$.
We fit the spectrum with the absorbed power law model using the
C-statistics  and obtained the photon index
$\Gamma=1.1^{+0.9}_{-0.8}$ and hydrogen column density $N_{\rm H} =
5.2^{+3.3}_{-2.4}\times 10^{22}$ cm$^{-2}$. The unabsorbed 0.5-10
keV flux is $F_X=5.4^{+2.0}_{-1.3}\times 10^{-12}$ erg cm$^{-2}$
s$^{-1}$.

 The combined {\sl Swift}-ISGRI spectrum
can be fitted with a single power-law model with $\Gamma =
1.3^{+0.2}_{-0.2}$, $N_{\rm H} = 6.1^{+3.2}_{-2.2}\times
10^{22}$ cm$^{-2}$.
Based on the positional coincidence and spectral compatibility,
we conclude that we detected a {\sl Swift} counterpart of IGR J2018+4043;
we will call this source J2018 hereafter.

Because of the presence of the bright ($m_V=5.82$) O9V star HD193322
in the FOV, only W2 and M2 ultraviolet filters were used with the
{\sl Swift}/UVOT camera. The source was not detected in these
filters, the upper limits being not
constraining given the
strong ISM extinction in this direction (see \S3).

\subsection{VLA data analysis}

The radio image at 1.4 GHz was produced from VLA archival data
obtained in D configuration on 1996 Sep 19.
The data were processed with the MIRIAD software
following standard procedures. The source 1411+522 was used
as a primary flux density calibrator ($S_{\rm 1.4\,GHz} = 22.8$ Jy)
and 1924+334 as a phase calibrator.
The
synthesized beam is
$54''\times 53''$ (beamwidth at half power),
P.A. = 33$^\circ$, and the rms noise is
about 5 mJy/beam.

Figure 1 shows
a contour image of the region
around
J2018, at 1.4 GHz.
The
J2018 position is
within an extended radio feature, of about $5'$ in size,
$\approx 70''$
south-east of the apparent position of radio intensity peak.
The total flux density of the radio feature is
about
480 mJy.
The resolution in D configuration is too low to resolve
a possible point source within the structured extended feature.
Overall, we cannot rule out the possibilty that the radio source
is a counterpart of J2018,
although the connection
is not very certain yet.

\subsection{Optical and infrared data analysis}

\begin{figure*}
\begin{center}
\includegraphics[width=0.6\textwidth,angle=270]{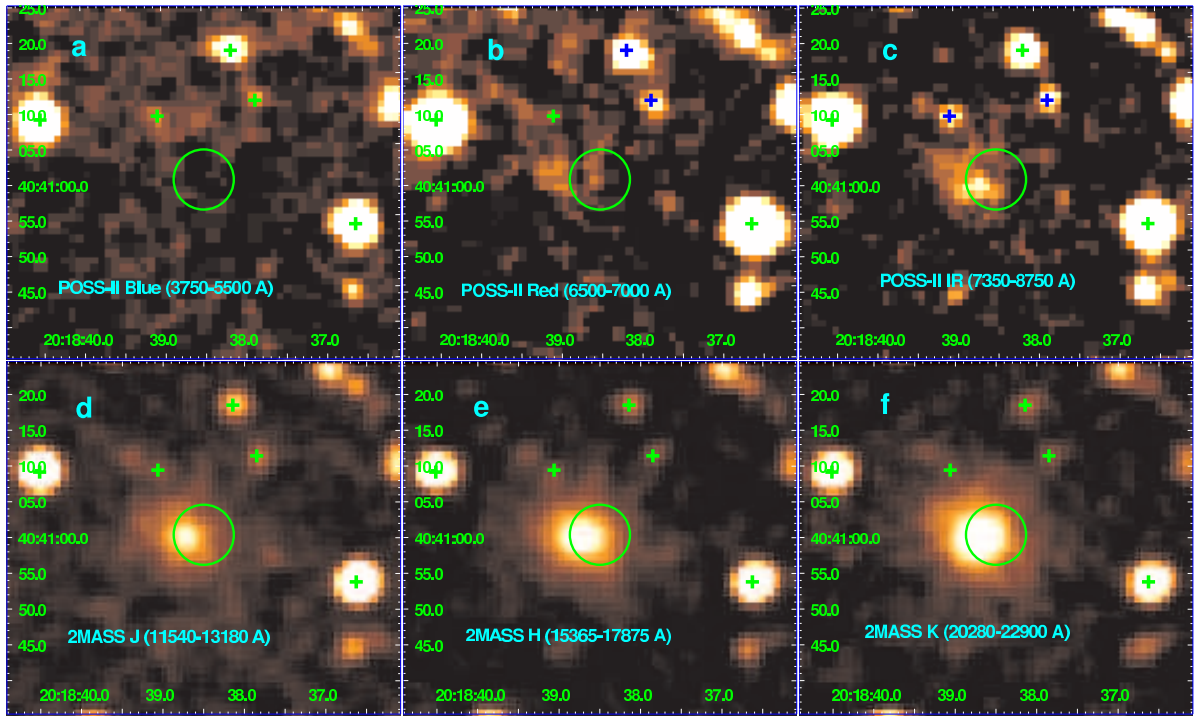}
\caption{ Archival
POSS-II
({\em top}) and 2MASS
({\em bottom})
images
of the J2018 field.
The {\sl Swift} \xrt\ position is indicated with
the 4\farcs2 radius
error circle.
The green and blue crosses denote point-like
sources from the GSC 2.3 catalog.
} \label{fig2}
\end{center}
\end{figure*}

Having the position of J2018 well constrained by the \sw\
observations, we looked for counterparts of the X-ray source in
archival optical and IR data. We found a likely counterpart, with
the flux of 88$\pm$9 mJy at 8.28 $\mu$m, in the 6th MSX catalog
\citep{egan} at $\alpha$ = 20$^{\rm h}$18$^{\rm m}$39$\fs$1, $\delta
= +40^\circ 40'56''$ (about $2''$ positional accuracy). We also
found an extended NIR source centered at $\alpha$ = 20$^{\rm
h}$18$^{\rm m}$38$\fs$73, $\delta$ = +40$^\circ$41$'$00$\farcs$1
(2\farcs6 from the {\sl Swift} position) in the 2MASS XSC catalog
\citep{2mass}, with magnitudes $J=13.0$, $H=11.5$, $K_s=10.7$ (Fig.\
2). These magnitudes are brighter than reported by Kennea et al.\
(2006) because those authors did not account for source extension.

%
The optical images,
taken from the POSS-II survey \citep{poss2}, are also
shown in Figure \ref{fig2}.
A counterpart to
J2018 can be seen in
the POSS-Red (5878-7121 \AA) and POSS-IR (6939-9030 \AA) filters
at the same position as the 2MASS source, and with a similar spatial extent.

We estimated the fluxes of the extended optical source, using the
magnitudes of two nearby stars from the GSC 2.3 catalog
\citep{gsc2}, marked with blue crosses in panels b and c of Figure
2. We estimated $m_{\rm Red} = 18.1$ and $m_{\rm IR} = 17.3$ via
aperture count-rate calculations. The absence of the source in the
POSS-II Blue image yields an upper limit $m_{\rm Blue}>22.5$.
The aperture boundaries were set at the count rate
corresponding to half a difference between the brightest source
pixel value and the local background level. Two reference stars in
each band were used to estimate the zero-point and the count-rate
vs.\ magnitude dependence. The same local background was subtracted
from all the count rates, normalized by the aperture size.
To convert the magnitudes into fluxes, we used the $m_{\rm AB}$
and POSS-II magnitudes of the standard secondary
spectrophotometric objects HZ\,4 and G191B2B \citep{oke}
and obtained
$F_{\rm Blue} < 2.5\mu$Jy, $F_{\rm Red} = 0.27$ mJy, $F_{\rm IR}$ =
0.76 mJy for the optical source. We estimate the error of the
cross-calibration, about 20\%, to be mainly due to the extended nature of
the source.

\section{Discussion}

The {\sl Swift}-ISGRI spectrum of J2018 shows that it is a
nonthermal, highly obscured source. The hydrogen column density
estimated from the {\sl Swift} data (\S2.2) looks too high for an
object in the \gc\ SNR ($d\approx 1.5$ kpc, expected $N_{\rm H} \sim
1$--$3\times 10^{21}$ cm$^{-2}$; Landecker et al. 1980). J2018 is
likely a more distant object, not associated with \gc.

A number of ISGRI sources have been identified as Galactic X-ray
binaries, mostly HMXBs (e.g., Filliarte \& Chaty 2004; Tomsick et
al.\ 2006, and references therein). The J2018's X-ray spectrum
(\S2.2) and luminosity, $L_X\sim 3\times 10^{35}(d/{\rm
10\,kpc})^2$ ergs s$^{-1}$ in 1--80 keV range, as well as the
magnitudes of the IR-optical counterpart (\S2.4) do not contradict
such an interpretation. The apparent $\sim$$10\arcsec$ extension
of the optical and 2MASS counterparts could be associated with a
powerful wind of an HMXB's O or B companion. Alternatively, the
extended IR-optical source could be interpreted as a galaxy that
hosts an AGN detected in X-rays.

An extrapolation of the X-ray spectrum of J2018 into the EGRET range
reaches the flux level of GEV J2020+4023, which suggests that these
sources could be X-ray and $\gamma$-ray counterparts of the same
object. An extensive dedicated search for a counterpart to 3EG
J2020+4017 with \chan\ and the Green Bank radio telescope in a close
vicinity of its nominal position failed to show any clear candidate
object (Becker et al.\ 2004; Weisskopf et al.\ 2006). However, J2018
is located outside the fields of these searches.

The position of J2018 is 30\farcm6 (35\farcm5)
 offset from the nominal position of GEV J2020+4023 (3EG
 J2020+4017), exceeding the $8'$ ($9\farcm6$) 95\% localization accuracy in those
catalogs (Lamb \& Macomb 1997; Hartman et al.\ 1999). However,
such offsets are not unusual among EGRET sources.
The position accuracy of relatively faint EGRET sources in the
Galactic plane was estimated by Gehrels et al.\ (2001) as
$\sim$$1\degmark$. Moreover, even the very bright Vela and Geminga
pulsars are located outside their EGRET 99\% confidence contours
(see Hartman et al.\ 1999). 3EG J2020+4017 is a bright source, but
it is located in a very crowded region of the plane, with enhanced,
highly inhomogeneous diffuse emission from the \gc\ SNR and nearby
molecular clouds, and at least two other EGRET sources (3EG
J2021+3716 and J2016+3657) within the 3$\degmark$ vicinity, which
may have affected the source localization. In such an environment,
systematic uncertainties can  increase the localization error up to
$\sim$0\fdg5. Thus, despite the positional discrepancy, we believe
that one can consider the possibility that {\sl Swift}-ISGRI and
EGRET detected the same source. Future {\sl GLAST} (e.g., Gehrels \&
Michelson 1999)
 observation could resolve the issue.

\begin{figure*}
\begin{center}
\includegraphics[width=0.95\textwidth,angle=0]{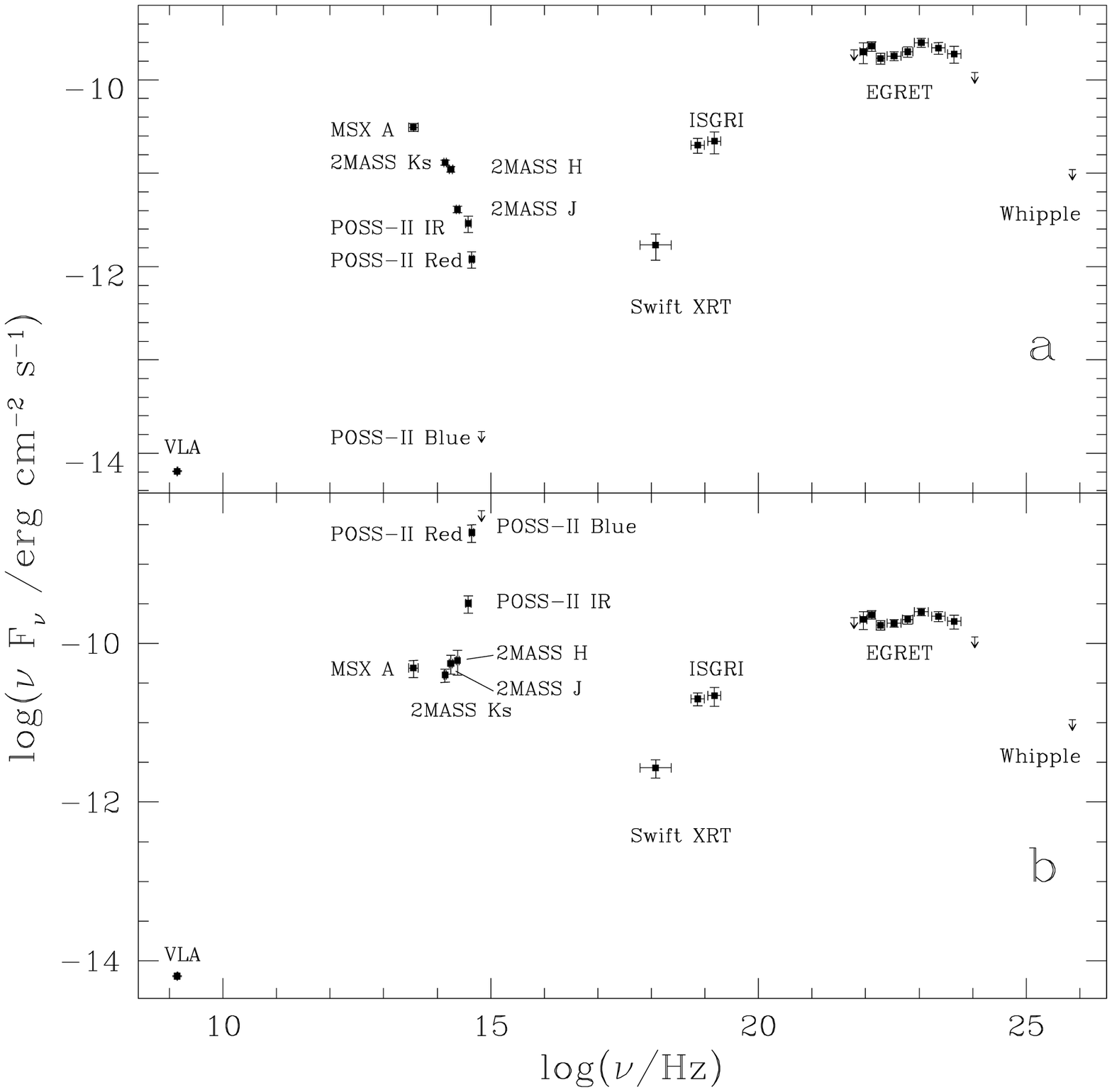}
\caption{
Observed ({\em top}) and extinction-corrected ({\em bottom})
multiwavelength spectra of J2018. The NIR-optical and soft X-ray
points in the bottom panel are corrected for absorption assuming
$A_V=9.8$ (see \S3) and $N_{\rm H}=6.1\times 10^{22}$ cm$^{-2}$,
respectively. Note that the observational data are
noncontemporaneous. The EGRET data points for 3EG\,J2020+4017 are
included, although the association with J2018  is not firmly
established (see \S3).} \label{fig3}
\end{center}
\end{figure*}

It is also worth mentioning the $\sim 3\,\sigma$ detection of a
possible Fe K$_\alpha$ feature at 6.2 keV in the {\sl RXTE} spectrum
of \gc\ (Bykov et al.\ 2004). The feature, which could be attributed
to a putative hidden accreting source (e.g., an AGN), does not
appear in the {\sl ASCA} spectra of the \gc\ regions observed by
this satellite, which did not include the J2018 position, while this
position was well within the {\sl RXTE} collimator field.


To investigate the nature of J2018, we constructed its spectral
energy distribution (SED; Fig.\ \ref{fig3}) using the \isgri\ and
\swift\ observations
 and the archival
data described above. It should be noted that the data are
noncontemporaneous, which is a point of some concern because the
source could be variable. The Galactic absorption ($A_V\approx 9.8$)
was accounted for in the unabsorbed NIR-optical spectrum using the
reddening maps by \citet{schleg}\footnote{See also
http://irsa.ipac.caltech.edu/applications/DUST/}. Being rather
accurate for high latitude sources, the method often overestimates
the reddening by a factor of up to 1.5 for strongly reddenned
objects (Arce \& Goodman 1999). The optical emission is apparently
extended, but with the data available it is not possible to
disentangle the core component, corresponding to the point source
detected  by {\sl Swift}, from the extended halo. Therefore, the
optical points can be lower than shown in Figure 3, and the
extinction corrected (with $A_V= 9.8$) optical fluxes in Figure
\ref{fig3} should be regarded as upper limits, because of a possible
overestimation (up to 3 mag) in the $A_V$ used.

The hypothesis that the 1.4\,GHz VLA excess is due to a point-like
source at the \swift\ source position (e.g., an AGN) projected onto
an extended radio source
is not in conflict with the data available. Therefore, we conditionally include the
VLA point, as well as the EGRET data, in Figure 3. Within a more
conservative approach, they should be regarded as
upper limits.

The radio through hard X-ray SED
 in Figure \ref{fig3} is consistent with that expected
for a nearby Seyfert AGN or a radio-galaxy (e.g., Dermer \& Gehrels
1995). For instance, at $z=0.02$ the source luminosity below 100 keV
would be about $10^{43}$ ergs s$^{-1}$.

  If GEV J2020+4023 is indeed associated with J2018, then the radio
through $\gamma$-ray source could be a blazar. Given the $A_V$
uncertainties discussed above, the multiwavelength spectrum
resembles the two-peak synchrotron self-Compton  SED of S5 0716+714,
a blazar studied by Foschini \etal (2006). Simultaneous
multiwavelength observations are required, however, to verify the
blazar hypothesis and understand the processes responsible for such
a spectrum. On the other hand, the leptonic jet model for a
microquasar, such as discussed by Dermer \& B\"{o}ttcher (2006) for
LS 5039, also provides a SED generally consistent with that given in
Figure 3.

NIR and optical spectroscopy could distinguish between the two
interpretations by measuring the redshift of the galaxy that hosts
the putative AGN. Also, microquasars are usually variable on minutes
scale, while AGNs are variable on longer time scales making a
variability study conclusive.

The analysis given above allows us to conclude that the properties of the
source discovered by \sw\ are consistent with those of
\igr. However, that does not exclude a sizable extended hard X-ray
flux from the possible interaction of the \gc\ SNR with stellar
winds as discussed by Bykov et al.\ (2004). The extended VLA feature
seen in Figure 1 can also be attributed to the SNR-wind interaction.
High-resolution observations with \chan\ and \xmm, as well as
simultaneous multiwavelength studies, are needed to separate the
components and to make a firm conclusion on the nature of IGR J2018+4043.

\acknowledgements
We thank the anonymous referee whose useful remarks helped us
to improve the presentation of the results.
Support for this work was provided by the NASA through grant NAG5-10865,
by RBRF grants 06-02-16844, 04-02-16595, and by France-Argentina
ECOS A04U03.

\end{document}